\documentclass[preprintnumbers,secnumarabic,amsmath,amssymb,nofootinbib,floatfix,superscriptaddress,english,12pt]{article}
\usepackage{array}
\usepackage{graphicx}
\usepackage{amssymb}
\usepackage[english]{babel}
\usepackage{amsmath}
\usepackage{multirow}
\usepackage{prettyref}
\usepackage{babel}
\usepackage{units}
\usepackage[latin1]{inputenc}
\usepackage{amsfonts}
\usepackage{amssymb}
\usepackage{babel}
\usepackage{color}
\usepackage{appendix}
\usepackage{cite}
\def\@fmsl@sh#1#2#3{\m@th\ooalign{$\hfil#1\mkern#2/\hfil$\crcr$#1#3$}}
 \def\eq#1\en{\begin{equation}#1\end{equation}}
\def\s[#1,#2]{[#1\stackrel{\star}{,}#2]}
\def\sx[#1,#2]{[#1\stackrel{\star_{x}}{,}#2]}

\newcommand{\ba}{\begin{eqnarray}}
\newcommand{\ea}{\end{eqnarray}}

\newcommand{\be}{\begin{eqnarray}}
\newcommand{\ee}{\end{eqnarray}}

\newcommand{\bc}{\hat b^\dagger}
\renewcommand{\d}{\mbox{${\rm d}$}} 
\newcommand{\lp}{\ell_{\rm p}}

\newcommand{\gn}{G_{\rm N}}
\newcommand{\rs}{R_{\rm s}}

\newcommand{\Rh}{R_{\rm H}}

\newcommand{\mn}{{\mu\nu}}

\newcommand{\nc}{\newcommand}
\nc{\beq}{\begin{equation}}
\nc{\eeq}{\end{equation}}
\nc{\beqa}{\begin{eqnarray}}
\nc{\eeqa}{\end{eqnarray}}

\def\bc{\begin{center}}
\def\ec{\end{center}}

\def\to{\rightarrow}

\def\gsim{\mathrel{\mathpalette\atversim>}}

\def\bc{\begin{center}}
\def\ec{\end{center}}

\def\gsim{\mathrel{\rlap{\lower4pt\hbox{\hskip1pt$\sim$}}

    \raise1pt\hbox{$>$}}}       

\def\gsim{\mathrel{\rlap{\lower4pt\hbox{\hskip1pt$\sim$}}
    \raise1pt\hbox{$>$}}}       

\begin{document}
\makeatletter
\def\fmslash{\@ifnextchar[{\fmsl@sh}{\fmsl@sh[0mu]}}
\def\fmsl@sh[#1]#2{%
  \mathchoice
    {\@fmsl@sh\displaystyle{#1}{#2}}%
    {\@fmsl@sh\textstyle{#1}{#2}}%
    {\@fmsl@sh\scriptstyle{#1}{#2}}%
    {\@fmsl@sh\scriptscriptstyle{#1}{#2}}}
\def\@fmsl@sh#1#2#3{\m@th\ooalign{$\hfil#1\mkern#2/\hfil$\crcr$#1#3$}}
\makeatother

\thispagestyle{empty}
\begin{titlepage}
\boldmath
\begin{center}
  \Large {\bf    Quantum Gravitational Corrections to a Star Metric and the Black Hole Limit}
    \end{center}
\unboldmath
\vspace{0.2cm}
\begin{center}
{  {\large Xavier Calmet}$^a$\footnote{x.calmet@sussex.ac.uk}},
{  {\large Roberto Casadio}$^{b,c}$\footnote{casadio@bo.infn.it}}
{and}
{  {\large Folkert Kuipers}$^a$\footnote{f.kuipers@sussex.ac.uk}}
 \end{center}
\begin{center}
{\sl $^a$Department of Physics and Astronomy, 
University of Sussex, Brighton, BN1 9QH, United Kingdom
}
\\
{\sl $^b$Dipartimento di Fisica e Astronomia, Universit\`a di Bologna, via Irnerio 46, I-40126 Bologna, Italy
}
\\
{\sl $^c$I.N.F.N., Sezione di Bologna, IS - FLAG, via B.~Pichat~6/2, I-40127 Bologna, Italy
}
\end{center}
\vspace{5cm}
\begin{abstract}
\noindent
In this paper we consider the full set of quantum gravitational corrections to a star metric to second order in curvature. As we use an effective field theoretical approach, these corrections  apply to any model of quantum gravity that is based on general coordinate invariance. We then discuss the black hole limit and identify an interesting phenomenon which could shed some light on the nature of astrophysical black holes: while star metrics receive corrections at second order in curvature, vacuum solutions such as black hole metrics do not. What happens to these corrections when a star collapses? 
\end{abstract}  
\vspace{5cm}
\end{titlepage}



\newpage
\section{Introduction}
\setcounter{equation}{0}
\label{s:intro}
Since the seminal work of Weinberg in 1979~\cite{Weinberg}, much progress has been made in quantum gravity using
effective field theory methods~\cite{Bar1984,Bar1985,Bar1987,Bar1990,Buchbinder:1992rb,Donoghue:1994dn,Calmet:2018elv}.
While finding a consistent theory of quantum gravity valid at all energy scales remains an elusive goal, effective field theory
methods can be applied at energies below the Planck mass which might be all that is ever needed as physics is an empirical
science.
This approach enables calculations in quantum gravity which are model independent,
see e.g.~\cite{X1,Donoghue,Bjerrum-Bohr,Calmet:2018uub,Calmet:2018qwg,Calmet:2016fsr,Calmet:2017rxl,Calmet:2015dpa,
Calmet:2018GravRadiation,CX1,CX2,CX3,CX4,CX5,Calmet:2014dea,Calmet:2011ta,Calmet:2008dg}.  The model independence only applies to models that assume that general coordinate invariance is also the correct symmetry of quantum gravity. Obviously in fundamental models with e.g. Lorentz violation, the effective field theory could be different.
One of the important results recently obtained is that there are no quantum gravitational corrections to vacuum solutions of general
relativity~\cite{Calmet:2018elv} to second order in curvature.
This in particular applies to eternal black hole metrics which are static vacuum solutions \cite{X1}.
On the other hand, real astrophysical black holes are clearly not in vacuum and they undergo a time evolution
as they are formed out of some time dependent astrophysical process such as during the collapse of a heavy star. 
\par
Understanding the transition from a star to a black hole state could help to understand the nature of astrophysical black holes better.
The aim of this work is to do a first step in that direction by calculating quantum gravitational corrections to the metric of a star
in stable equilibrium, as described by the Tolman-Oppenheimer-Volkoff equation.
In general relativity, the metric outside non-rotating black holes and stars is given in both cases by the vacuum
Schwarzschild solution.
Our aim is to compare the quantum gravitational corrections to a star metric and black hole metric as seen by an observer
who is far away from both objects.
While it is known that in the black hole case there are no corrections to the metric at second order in curvature,
we will show that there is a correction at this order in the case of a star.
 This phenomenon is intriguing as a distant observer could in principle differentiate a star that is collapsing from an eternal black hole (i.e. a vacuum solution) by measuring the correction at order $\gn^2$ to Newton's potential. The collapsing star would have a potential that deviates from $1/r$ by corrections of order $\gn^2$ while the black hole vacuum solution does not have such corrections.
\par
We then  consider the limit when the mass and the radius of the star are taken towards respectively the
Planck mass and the Planck length and discuss whether the metric obtained in that limit could be
used to describe the metric of a quantum black hole, i.e.~the lightest black holes that could
have masses of the order of the Planck mass and a Schwarzschild radius of the order of the Planck length.
We argue that as quantum black holes cannot be described as a classical vacuum,
the quantum corrected star metric should be a better model for the metric of a quantum black hole than the
Schwarzschild vacuum solution.
\par
This paper is organized as follows. In Section~\ref{S:QC}, we introduce the effective quantum gravitational action
and calculate the leading order corrections to the metric for a homogeneous isotropic star. In Section~\ref{S:div},
we discuss the validity of our results close to the surface of the star.
In Section~\ref{s:Smetric}, we discuss the differences with an eternal Schwarzschild black hole metric and argue
that quantum black holes might be better described by the star metric.
Finally, we conclude with some outlooks in Section~\ref{s:outlook}.
\section{Quantum corrections to a star metric}
\label{S:QC}
\setcounter{equation}{0}
Aim of this section is to calculate the leading order quantum gravitational corrections to the metric of a stable star
satisfying the Tolman-Oppenheimer-Volkoff equation.
This investigation was started in~\cite{X1}, but that paper only considered the contribution of the term $R\, \log \Box\, R$.
Here we consider the full set of corrections at second order in curvature.
We also take this opportunity to fix a calculational mistake in~\cite{X1}. 
\par
We work within the framework of the effective quantum gravitational action given
by~\cite{Weinberg,Bar1984,Bar1985,Bar1987,Bar1990,Buchbinder:1992rb,Donoghue:1994dn,Calmet:2018elv}  
\be
\Gamma[g] = \Gamma_{\rm L}[g] + \Gamma_{\rm NL}[g]
\ ,
\ee
where the local part of the action is given by~\footnote{In this paper we work in the $(+---)$ signature
and use the convention where the Riemann tensor is defined by
$R^{\rho}_{~\sigma\mu\nu} = \partial_{\mu} \Gamma^{\rho}_{\nu\sigma} - ...$
and the Ricci tensor by $R_{\mu\nu}=R^{\lambda}_{~\mu\lambda\nu}$}
\be
\Gamma_{\rm L}
=
\int \d^4 x\, \sqrt{g} 
\left[ \frac{R}{16\,\pi\, \gn}
+
c_1(\mu)\, R^2
+ c_2(\mu)\, R_\mn\, R^\mn
+ c_3(\mu)\, R_{\mu\nu\alpha\beta}\,
R^{\mu\nu\alpha\beta}
\right]
\ee
and the non-local part of the action by
\be
\Gamma_{\rm NL}
=
-\int \d^4 x\, \sqrt{g}
\left[
\alpha\, R\, \ln{\!\left(\frac{\Box}{\mu^2}\right)}R
+
\beta\, R_\mn\, \ln\!{\left(\frac{\Box}{\mu^2}\right)}
R^\mn
+
\gamma\, R_{\mu\nu\alpha\beta}\, \ln\!{\left(\frac{\Box}{\mu^2}\right)}
R^{\mu\nu\alpha\beta}
\right]
\ .
\label{eq:NLterms}
\ee
This effective action is obtained by integrating out the fluctuations of the graviton and potentially
other massless matter fields.
While the Wilson coefficients of the local part of the action are not calculable from first principles
as we do not specify the ultra-violet theory of quantum gravity, those of the non-local part are
calculable and model independent quantum gravitational predictions.
We reproduce these coefficients, which have  been derived by many different authors,
see e.g.~\cite{Birrell:1982ix,Mirzabekian:1993nz,Donoghue:1994dn,Elizalde:1995tx,
Han:2004wt,Mirzabekian:1998ha,Bar1984,Bar1985,Kallosh:1978wt,Donoghue}, in Table~\ref{coeff1}.
\begin{table}
\center
\begin{tabular}{| c | c | c | c |}
\hline
 & $ \alpha $ & $\beta$ & $\gamma$  \\
 \hline
 \text{Scalar} & $ 5(6\xi-1)^2$ & $-2 $ & $2$     \\
 \hline
 \text{Fermion} & $-5$ & $8$ & $7 $ \\
 \hline
 \text{Vector} & $-50$ & $176$ & $-26$ \\
 \hline
 \text{Graviton} & $250$ & $-244$ & $424$\\
 \hline
\end{tabular}
\caption{Non-local Wilson coefficients for different fields. 
All numbers should be divided by $11520\pi^2$. Here, $\xi$ denotes the value of the non-minimal coupling for a scalar theory. 
All these coefficients including those for the graviton are gauge invariant.
It is well known that one needs to be careful with the graviton self-interaction diagrams and that the coefficients
$\alpha$ and $\beta$ can be gauge dependent, see~\cite{Kallosh:1978wt}, if the effective action is defined in a naive way.
For example, the numbers $\alpha= 430/(11520\pi^2)$ and $\beta=-1444/(11520\pi^2)$ for the graviton quoted
in~\cite{Donoghue} are obtained using the Feynman gauge.
However, there is a well-established procedure to derive a unique effective action which leads to gauge independent
results~\cite{Bar1984,Bar1985}.
Here we are quoting the values of $\alpha$ and $\beta$ for the graviton obtained using this formalism as it guaranties
the gauge independence of observables.}
\label{coeff1}
\end{table}
\par
The equations of motion obtained from varying the effective action which respect to the metric are given by
\be
G_\mn + 16\,\pi\,\gn \left( H_\mn^{\rm L} + H_\mn^{\rm NL} \right)
=
0
\ ,
\ee
where
\be
G_\mn
=
R_\mn - \frac{1}{2}\, R\, g_\mn
\ee
is the usual Einstein tensor.
The local part of the equation of motion is given by
\begin{align}
H_\mn^{\rm L} 
=
&\,  
\bar{c}_1
\left( 2\, R\, R_\mn - \frac{1}{2}\, g_\mn\, R^2 + 2\, g_\mn\, \Box R - 2 \nabla_\mu \nabla_\nu R\right) 
\label{eq:EQMLoc}
\\
&\,
+\bar{c}_2
\left( 2\, R_{~\mu}^\alpha\, R_{\nu\alpha} - \frac{1}{2}\, g_\mn\, R_{\alpha\beta}\, R^{\alpha\beta}
+ \Box R_\mn + \frac{1}{2}\, g_\mn\, \Box R
- \nabla_\alpha \nabla_\mu R_{~\nu}^\alpha
- \nabla_\alpha \nabla_\nu R_{~\mu}^\alpha \right)
\ ,
\nonumber
\end{align}
with $\bar{c}_1 = c_1 - c_3$ and $\bar{c}_2 = c_2 + 4\, c_3$.
Finally, the non-local part reads
\begin{align}
H_\mn^{\rm NL} 
=
&\,
 - 2\,\alpha
 \left( R_\mn - \frac{1}{4}\, g_\mn\, R
 + g_\mn\, \Box
 - \nabla_\mu \nabla_\nu \right)
 \ln\left(\frac{\Box}{\mu^2}\right)\, R
 \nonumber
 \\
&\, 
- \beta
\bigg( 2\, \delta_{(\mu}^\alpha\, R_{\nu)\beta}
- \frac{1}{2}\, g_\mn\, R_{~\beta}^\alpha
+ \delta_{\mu}^\alpha\, g_{\nu\beta}\, \Box
+ g_\mn\, \nabla^\alpha \nabla_\beta  \nonumber  \\ \nonumber & 
- \delta_\mu^\alpha\, \nabla_\beta \nabla_\nu
- \delta_\nu^\alpha\, \nabla_\beta \nabla_\mu 
\bigg)
\ln\left(\frac{\Box}{\mu^2}\right)\, R_{~\alpha}^\beta &
\nonumber
\\
&\,
- 2 \,\gamma
\left(
\delta_{(\mu}^\alpha\, R_{\nu)~\sigma\tau}^{~\beta}
- \frac{1}{4}\, g_\mn\, R^{\alpha\beta}_{~~\sigma\tau}
+\left( \delta_\mu^\alpha\, g_{\nu\sigma} + \delta_\nu^\alpha\, g_{\mu\sigma} \right)
\nabla^\beta \nabla_\tau \right)
\ln\left(\frac{\Box}{\mu^2}\right)\, R_{\alpha\beta}^{~~\sigma\tau}
\ .
\end{align}
Note that the variation of the $\ln \Box$ term yields terms of higher order in curvature
and can thus safely be ignored at second order in curvature. 
\par
We consider a stationary homogeneous and isotropic star with density 
\be
\rho(r)
=
\rho_0\,
\Theta(\rs-r)
=
\begin{cases}
\rho_0
&
{\rm if}\ r<\rs
\\
0
&
{\rm if}\ r>\rs
\ ,
\end{cases}
\label{rho0}
\ee
where $\rho_0>0$ is a constant and $\Theta(x)$ is Heaviside's step function.
The solution to the Einstein equation inside this star (for $r\le\rs$) is the well-known interior Schwarzschild
metric~\cite{schw16,Wald}
\begin{align}
\label{eq:IntSchw}
\d s^2
&=
\left( 3\, \sqrt{1 - \frac{2\, \gn\, M}{\rs}} - \sqrt{1 - \frac{2\, \gn\, M \,r^2}{\rs^3}} \right)^2 
\frac{\d t^2}{4} 
-
\left(1 - \frac{2\, \gn\, M r^2}{\rs^3}\right)^{-1}\d r^2
-
r^2\,\d\Omega^2
\nonumber
\\
&\equiv
g_{\mu\nu}^{\rm int}\,\d x^\mu\,\d x^\nu
\ ,
\end{align}
where
\be
M
=
4\,\pi
\int_0^{\rs}
\rho\,r^2\,\d r
=
\frac{4\,\pi}{3}\,\rs^3\,\rho_0
\ee
is the total Misner-Sharp mass of the source. 
The corresponding pressure is given by
\be\label{eq:StarPress}
P(r)
= 
\rho_0\, \frac{ \sqrt{1 - \frac{2\, \gn\, M}{\rs}}
-\sqrt{1 - \frac{2\, \gn\, M \,r^2}{\rs^3}}}
{ \sqrt{1 - \frac{2\, \gn\, M\, r^2}{\rs^3}}
- 3\, \sqrt{1 - \frac{2\, \gn\, M}{\rs}}} 
\ 
= \mathcal{O}(\gn)
\ ,
\ee
and is of order $\gn$ in agreement with the fact that the pressure does not gravitate in
Newtonian physics.
Of course, the metric outside the star (for $r>\rs$) is the usual vacuum Schwarzschild
metric~\cite{schw1916,Wald}
\be
\label{eq:ExtSchw}
\d s^2 =
\left(1 - \frac{2\,\gn\, M}{r}\right) \d t^2
-
\left(1 - \frac{2\,\gn\, M}{r}\right)^{-1}
\d r^2
- r^2\,\d\Omega^2
\equiv
g_{\mu\nu}^{\rm ext}\,\d x^\mu\,\d x^\nu
\ ,
\ee
from which one can see that $M$ is also the Arnowitt-Deser-Misner (ADM) mass~\cite{adm} of the system.
\par
We now perturb the above metrics,
\be
\tilde{g}_\mn = g_\mn + g_\mn^{\rm q}
\ ,
\ee
and take the perturbation $g_\mn^{\rm q}$ to be $\mathcal{O}(\gn)$.
The equations of motion then become
\be
\label{eq:EQMLin}
G_\mn^{\rm L}[g^{\rm q}]
+16\,\pi\,\gn
\left( H_\mn^{\rm L}[g] + H_\mn^{\rm NL}[g] \right)
=
0
\ ,
\ee
where the linearised Einstein tensor is given by
\begin{align}
2\,G_\mn^{\rm L}
=
&\,
\Box g_\mn^{\rm q} - g_\mn \,\Box g^{\rm q}
+ \nabla_\mu \nabla_\nu g^{\rm q} 
+ 2\,R^{\alpha~\beta}_{~\mu~\nu} \,g_{\alpha\beta}^{\rm q} 
- \nabla_\mu \nabla^\beta g_{\nu\beta}^{\rm q} 
- \nabla_\nu \nabla^\beta g_{\mu\beta}^{\rm q}
\nonumber
\\
&\,
+ g_\mn \,\nabla^\alpha \nabla^\beta g_{\alpha\beta}^{\rm q}
\ .
\end{align}
\par
 We first calculate solutions to equation~(\ref{eq:EQMLin}) due to the local corrections. Outside the star, where the unperturbed metric equals the Schwarzschild vacuum solution~\eqref{eq:ExtSchw} with $R=R_\mn=0$, these corrections are trivially $0$. Inside the star this is not the case. However, these corrections turn out to be $\mathcal{O}(\gn^3)$, and thus sub-leading.
Therefore the local part in the equations of motion~(\ref{eq:EQMLoc}) does not contribute.
\par
 In order to calculate corrections due to the non-local corrections of the equation of motion~(\ref{eq:EQMLoc}) we use the fact that the Ricci Scalar, Ricci tensor and Riemann tensor are all
$\mathcal{O}(\gn)$. We thus obtain
\begin{align}
\frac{G_\mn^{\rm L}}{16\,\pi\,\gn} 
=
&\,
2\,\alpha 
\left( g_\mn\, \Box - \nabla_\mu \nabla_\nu \right)
\ln\!{\left(\frac{\Box}{\mu^2}\right)} R
\nonumber
\\
&\, 
+ \beta 
\left(
\delta_{\mu}^\alpha\, g_{\nu\beta}\,\Box
+ g_\mn \,\nabla^\alpha \nabla_\beta
- \delta_\mu^\alpha\, \nabla_\beta \nabla_\nu
- \delta_\nu^\alpha\, \nabla_\beta \nabla_\mu \right)
\ln\!{\left(\frac{\Box}{\mu^2}\right)}
R_{~\alpha}^\beta
\nonumber
\\
&\,
+ 2 \gamma 
\left( \delta_\mu^\alpha\, g_{\nu\sigma}
+ \delta_\nu^\alpha\, g_{\mu\sigma} \right)
\nabla^\beta \nabla_\tau 
\ln\!{\left(\frac{\Box}{\mu^2}\right)}
R_{\alpha\beta}^{~~\sigma\tau}
+ \mathcal{O}(\gn^3)
\ .
\label{eq:EQMNonLoc}
\end{align}
We will solve this equation perturbatively in $\gn$.
We use Einstein equations to rewrite the Ricci scalar and tensor in terms of the energy-momentum
tensor of the source,
\begin{align}
R &= - 8\, \pi\, \gn T
\\
R_\mn &= 8\, \pi\, \gn
\left( T_\mn - \frac{1}{2} \,g_\mn\, T \right)
\ ,
\end{align}
where, for a perfect isotropic fluid like our star, we have
\begin{align}
T &= \rho_0 + \mathcal{O}(\gn)
\\
T_\mn &= \delta_\mu^0 \,\delta_\nu^0 \,\rho_0
+
\mathcal{O}(\gn)
\ ,
\end{align}
where $\rho_0$ is the energy density.
\par
By applying the results from Appendix~\ref{A:PV} to the homogeneous distribution~\eqref{rho0},
we find
\be
8\, \pi\, \gn\,\ln\!{\left(\frac{\Box}{\mu^2}\right)}\,\rho
=
\frac{6 \gn M}{\rs^3}\, f(r) + \mathcal{O}(\gn^2)
\ ,
\ee
with
\begin{align}
	f(r) = 
	\begin{cases}
		-2\strut\displaystyle\left[ \gamma_E -1 + \ln\left( \mu \sqrt{\rs^2 - r^2} \right)\right]
		\quad
		&
		\textrm{if}\ r<\rs
		\ ,
		\\
		\\
		2\strut\displaystyle\frac{\rs}{r} - \ln \left( \frac{r + \rs}{r - \rs} \right)
		\quad
		&
		\textrm{if}\ r>\rs
		\ .
	\end{cases}
	\label{gr}
\end{align}
Note that the function $f$ in equation~\eqref{gr} is not defined at $r=\rs$.
In fact, one can verify that the results should be taken with some care in a small region around $\rs$, as we discuss
in more detail in Section~\ref{S:div}.
\par
Furthermore, we emphasize that equation (\ref{gr}) is the main source of the discrepancy between the results reported
here and those obtained in \cite{X1}, where the calculation was only done for $r>\rs$.
In equation~(31) of~\cite{X1} a factor of 2 is missing in front of the term $\rs/r$ and a factor of $-1$ is missing in front
of the $\log$ term.
\par
In order to obtain the contribution proportional to $\gamma$ in equation~(\ref{eq:EQMNonLoc}), we first rewrite
it in terms of those proportional to $\alpha$ and $\beta$ using the non-local Gauss-Bonnet theorem 
{\color{blue}~\cite{Calmet:2018elv,CovPertThII,CovPertThIII,CovPertThIV}},
which holds for the non-local part up to second order in curvature (hence $\mathcal{O}(\gn^2)$).
We then evaluate equation~(\ref{eq:EQMNonLoc}) using $\alpha' = \alpha - \gamma$ and
$\beta' = \beta + 4\,\gamma$.
We thus have to solve
\begin{align}
	G_\mn^{\rm L} 
	=
	&\,
	192\, \pi\, (\alpha -\gamma)\, \frac{\gn^2\, M}{\rs^3}
	\left(\nabla_\mu \nabla_\nu - g_\mn\, \Box \right) f(r)
	\nonumber
	\\
	&\, 
	+
	96\, \pi \,(\beta + 4\,\gamma)\, \frac{ \gn^2 \,M}{\rs^3}
	\left( \nabla_\mu \nabla_\nu - g_\mn\, \Box + \delta_\mu^0\, g_{\nu 0}\, \Box \right) f(r)
	+ \mathcal{O}(\gn^3)
	\ ,
\end{align}
where we used that
\begin{equation}
	\left( g_\mn\, \nabla^0 \nabla_0 - \delta_\mu^0\, \nabla_0\nabla_\nu - \delta_\nu^0\, \nabla_0\nabla_\mu \right)
	f(r)
	=
	\mathcal{O}(\gn)
	\ .
\end{equation}
We solve this equation, imposing the solution to be spherically symmetric and time independent.
In addition we fix the gauge freedom by setting $g_{\theta\theta}^{\rm q}=0$.
Doing so, we obtain the quantum corrections $g_\mn^{\rm q}=\delta g_\mn^{\rm ext}$ to the
Schwarzschild metric~\eqref{eq:ExtSchw} outside the star. The corrections are
given by~\footnote{Note that we take the metric with signature $(+---)$.
With signature $(-+++)$ the corrections obtain an extra minus sign.}
\begin{align}
\label{eq:MetricCorr}
	\delta g_{tt}^{\rm ext}
	&=
	(\alpha + \beta + 3\,\gamma)\,
	\frac{192 \,\pi\, \gn^2\, M}{\rs^3}
	\left[ 2\, \frac{\rs}{r} + \ln\! \left( \frac{r-\rs}{r+\rs} \right)
	\right] 
	+ \frac{C_1}{r}
	+ C_2
	+\mathcal{O}(\gn^3)
	\nonumber
	\\
	\delta g_{rr}^{\rm ext}
	&=
	(\alpha - \gamma) \,\frac{384\, \pi\, \gn^2\, M}{r \,(r^2 - \rs^2)}
	+\frac{C_1}{r}
	+ \mathcal{O}(\gn^3)
	\ ,
\end{align}
where $C_i$ are integration constants which must be set to zero, if we require asymptotic flatness, that is
$\lim_{r \rightarrow\infty}\delta g_\mn = \lim_{r \rightarrow \infty} r \,\delta g_\mn = 0$~\footnote{These
conditions ensure that we recover the classical weak field limit with ADM mass $M$ as $r\rightarrow\infty$,
which is the usual boundary condition for the classical Schwarzschild black hole.}.
\par
In a similar way, using the same gauge condition, one can find the corrections $g_\mn^{\rm q}=\delta g_\mn^{\rm int}$
to the metric~\eqref{eq:IntSchw} inside the star. These are given by
\begin{align}
\label{eq:MetricCorrInside}
	\delta g_{tt}^{\rm int}
	&=
	(\alpha + \beta + 3\,\gamma)\, \frac{192\, \pi\, \gn^2\, M}{\rs^3}\,
	\ln\! \left(\frac{\rs^2}{\rs^2 -r^2} \right)
	+ \frac{C_3}{r}
	+ C_4
	+ \mathcal{O}(\gn^3)
	\nonumber
	\\
	\delta g_{rr}^{\rm int}
	&=
	(\alpha - \gamma)\, \frac{384\, \pi\, \gn^2\, M\, r^2}{\rs^3\, (\rs^2 - r^2)}
	+ \frac{C_3}{r}
	+ \mathcal{O}(\gn^3)
	\ ,
\end{align}
where $C_i$ are integration constants, which we will set to $0$ by requiring regularity in the 
origin $r=0$.
\par
In the limit $r\to\rs$ we find that the corrections diverge, but it is easy to explain that these divergences
are generated, because we assumed a model for the star described by a discontinuous density at $r=\rs$,
which is not realistic for an astrophysical star.
This discontinuity leads to a discontinuity in the first derivative of the pressure~(\ref{eq:StarPress}),
in the second derivative of the $g_{tt}$ component and in the first derivative of the $g_{rr}$ component.
We thus do not expect that our star model and hence the quantum corrections apply to a real star
in a small region around $\rs$.
We shall discuss this observation in more details as well as how to cure these divergences in the next section.
\par
We can now consider our result in different limits. Far away from the star (for $r\gg\rs$), the leading behavior
of the metric corrections~(\ref{eq:MetricCorr}) is given by
\begin{align}\label{eq:MetricCorrLim}
	\delta g_{tt}^{\rm ext}
	&=
	- (\alpha + \beta + 3\,\gamma)\, \frac{128\, \pi\, \gn^2\, M}{r^3}
	+ \mathcal{O}(\gn^3)
	\nonumber
	\\
	\delta g_{rr}^{\rm ext}
	&=
	(\alpha - \gamma)\, \frac{384\, \pi\, \gn^2\, M}{r^3}
	+ \mathcal{O}(\gn^3),
\end{align}
whereas, to the same order in $\gn$, the corrections~\eqref{eq:MetricCorrInside} for the metric inside the star
far away from the star radius (for $r\ll\rs$) vanish,
\be
	\delta g_{tt}^{\rm int} = \delta g_{rr}^{\rm int} = \mathcal{O}(\gn^3)
	\ .
\ee
\par
It is important to realize that the correction to the components of a metric are gauge dependent.
As such components are not observables, this is not an issue.
For example, one could calculate the metric corrections in the harmonic gauge.
In this case one finds the asymptotic $r\gg\rs$ expressions
\begin{align}
	g_{tt}
	=
	&\,
	1 - \frac{2 \,\gn\, M}{r} 
	+ \frac{2\,\gn^2\, M^2}{r^2}
	- (\alpha + \beta + 3\,\gamma)\, \frac{128\, \pi\, \gn^2\, M}{r^3}
	+ \mathcal{O}(\gn^3)
	\nonumber
	\\
	g_{ti}
	=
	&\,
	0
	\nonumber
	\\
	g_{ij}
	=
	&\,
	- \delta_{i j} \left\{
	1 + \frac{2\, \gn M}{r} + \frac{\gn^2\, M^2}{r^2}
	- (2\,\alpha + \beta + 2\,\gamma)\, \frac{128\, \pi\, \gn^2\, M}{r^3}
	\left[\frac{1}{3} + \ln\!\left(\frac{C\,r}{\rs}\right)
	\right]
	\right\}
	\nonumber
	\\
	&\,
	 -\frac{x_i x_j}{r^2}
	 \left[ \frac{\gn^2\, M^2}{r^2} 
	 - (\alpha - \gamma)\, \frac{384\, \pi\, \gn^2\, M}{r^3}
	 + (2\,\alpha + \beta + 2\,\gamma)\, \frac{384\, \pi\, \gn^2\, M}{r^3}
	 \ln\!\left(\frac{C\,r}{\rs}\right)
	 \right]
	 \nonumber
	 \\
	 &\,
	 + \mathcal{O}(\gn^3)
	 \ ,
\end{align}
where $C$ is a dimensionless integration constant~\footnote{As in previous results, one obtains a couple more
integration constants, which can be set to $0$ by requiring that one recovers the classical weak field limit as $r\to \infty$.}.
We derived this result using the expression for the Schwarzschild metric outside a star in the harmonic gauge,
which can, for example, be found in~\cite{Weinberg2}.
Furthermore, we imposed the solutions to be spherically symmetric and time independent and imposed the harmonic
(De Donder) gauge condition instead of setting $\delta g_{\theta\theta}=0$.
\par
Taking the graviton values for $\alpha$, $\beta$ and $\gamma$ from~\cite{Donoghue},
one can set the scale $C/\rs=\mu\exp(-173/132)$, to recover the quantum correction due to the vacuum
polarization diagram found in~\cite{Bjerrum-Bohr}.
It should be emphasized that the graviton values for $\alpha$ and $\beta$ presented in~\cite{Donoghue}
are not gauge invariant~\cite{Kallosh:1978wt} and do not correspond to the values obtained when the unique
effective action formalism~\cite{Bar1984} is used, which are presented in Table~\ref{coeff1}.
The results in~\cite{Donoghue,Bjerrum-Bohr} are thus dependent on the gauge in which the effective action is obtained.
The results presented in this paper on the other hand do not suffer from this gauge dependence.
Naturally, both the results presented in this paper and those in~\cite{Donoghue,Bjerrum-Bohr} depend on the gauge
(that is, the reference frame) in which the field equations are solved.
This gauge dependence cannot be removed, as the metric components are not gauge invariant quantities. 
\par
Let us emphasize that the results presented in this section are interesting:
we have shown that although the metric outside an eternal static black hole and of a static star are given at the classical level
by the Schwarzschild solution, quantum gravity makes a difference between the two objects due to its non-local nature.
The star metric receives a quantum correction at second order in curvature, while there is no such correction for an eternal
black hole  \cite{Calmet:2018elv}.
A distant observer can in principle monitor the gravitational collapse of a star by studying the quantum gravitational
corrections to Newton's potential to second order in curvature.  This raises the question whether astrophysical black holes should really be described by metrics corresponding to vacuum solutions of General Relativity. Note that our argument does not rely on the limit $R_s \to 0$, but rather on a comparison of the initial state (e.g. collapsing star or star before it has even started to collapse) and the final state which is an eternal black hole.
\section{Divergence at the surface}
\label{S:div}
\setcounter{equation}{0}
The explicit calculation shown in Appendix~\ref{A:PV} makes it clear that the non-local function
$\ln\left(\frac{\Box}{\mu^2}\right)$ must be treated as a distribution in order to allow for the various
exchanges of limits and integrations.
This in turn implies that the functions $f$ upon which it can act must belong to a suitable set
of regular test functions. 
Clearly, the density profile~\eqref{rho0} does not satisfy this requirement, the Heaviside function
$\Theta$ being a distribution itself.
It therefore comes as no surprise that $\ln\left(\frac{\Box}{\mu^2}\right)\,\rho$ is not well defined around $r=\rs$,
unless the density~\eqref{rho0} is replaced with a function that falls to zero smoothly.
\par
It is important to remark that, although the density~\eqref{rho0} generating the classical Schwarz\-schild interior
metric~\eqref{eq:IntSchw} drops to zero within a vanishingly short length, it causes no issues in general relativity
despite the fact that the manifold is not smooth at the star surface.
Instead, it conjures with the non-local terms of the effective action~\eqref{eq:NLterms} to give rise to divergences.  The divergence thus purely arises due to inclusion of higher order derivatives of the metric, while the metric is only once continuously differentiable.
However, it is obvious that the density profile of any realistic matter distribution will
go to zero in a finite width $\epsilon>0$.
For instance, we could replace~\eqref{rho0} with the infinitely smooth 
\be
\rho(r)
=
\begin{cases}
	\rho_0\,\exp\left(\strut\displaystyle\frac{\epsilon^2}{\rs^2}-\frac{\epsilon^2}{\rs^2-r^2}\right)
	&
	{\rm for}\ 0\le r\le \rs
	\\
	0
	&
	{\rm for}\ \rs<r\ ,
\end{cases}
\ee
where we can safely assume that $\epsilon\gtrsim \lp$.
This implies that our solutions~\eqref{eq:MetricCorr} and~\eqref{eq:MetricCorrInside}
should only be considered outside a layer of thickness $\epsilon$ around $\rs$.
On the other hand, it is important to remark that the size of the corrections does not
depend on $\epsilon$ explicitly (only the region of space excluded in our results does). 
\par
In some more details, Eqs.~\eqref{eq:MetricCorr} and~\eqref{eq:MetricCorrInside}
contain divergences for $\epsilon\equiv |r-\rs|\to 0^+$,
namely 
\begin{align}
\delta g_{tt}^{\rm int}
&\simeq
-(\alpha+\beta+3\,\gamma)\, \frac{192\, \pi\, \gn^2\, M}{\rs^3}\,  \ln\!\left(\frac{2\,\epsilon}{\rs}\right) 
\nonumber
\\
\delta g_{tt}^{\rm ext}
&\simeq
(\alpha+\beta+3\gamma)\,\frac{192\, \pi\, \gn^2\, M}{\rs^3}
\left[2 +  \ln\!\left(\frac{\epsilon}{2\,\rs}\right) \right]
\nonumber
\\
\delta g_{rr}^{\rm int}
&\simeq
(\alpha -\gamma)\, \frac{192\, \pi\, \gn^2\, M}{\rs^3}
\left(\frac{\rs}{\epsilon} - \frac{3}{2} \right)
\nonumber
\\
\delta g_{rr}^{\rm ext}
&\simeq
(\alpha -\gamma)\, \frac{192\, \pi\, \gn^2\, M }{\rs^3}
\left(\frac{\rs}{\epsilon} - \frac{1}{2} \right)
\ ,
\end{align}
which appear in two forms, namely
\be
d_1
\sim
\frac{\gn^2\,M}{\rs^3}\,\ln\left(\frac{|r-\rs|}{r+\rs}\right)
\ ,
\ee
or
\be
d_2
\sim
\frac{\gn^2\,M}{r\left|r^2-\rs^2\right|}
\ .
\ee
Since we obtained the corrections in a ``weak'' field approximation, 
such terms should be small compared to the unperturbed metric coefficients, that is
\be
d_i
\lesssim
V
\sim
\frac{\gn\,M}{r}
\ .
\ee 
By recalling that $\gn=\lp^2$ in our units, this means that $d_1\ll V$ provided
\be
\frac{\lp^2}{\rs^2}\,
\ln\left(\frac{|r-\rs|}{\rs}\right)
\lesssim
1
\ee
and $d_2\ll V$ if
\be
\frac{\lp^2}{\rs\left|r-\rs\right|}
\lesssim
1
\ .
\ee
The above two conditions are clearly satisfied if $\epsilon\equiv |r-\rs|\lesssim \lp$, since
$\rs\gg \lp$ is the radius of a macroscopic matter source.
To illustrate this, one can derive numerical estimates on the size of $\epsilon$ for various
values of $\rs$.
In particular, we find for a typical neutron star with radius $\rs\simeq10\,\textrm{km}$,
that $\epsilon\gtrsim10^{-78}\,\rs$, while for objects of the order of the Planck length
$\rs \approx 10^{-35}\,\textrm{m}$, we find $\epsilon\approx\rs$.
As expected, our approximation fails  for sub-Planckian objects, and we must therefore
restrict our analysis to $M\gtrsim 1/\sqrt{G_N}=M_P$ where $M_P$ is the Planck scale. Moreover for Planck sized objects these restrictions are of major importance, and must be considered in any further analysis.
\section{Model for quantum black holes?}
\setcounter{equation}{0}
\label{s:Smetric}
While it is remarkable to be able to calculate model independent quantum gravitational corrections
to the metric of a star or vacuum solutions of general relativity, it is clear that these corrections are
tiny and probably of little empirical value from an astrophysical perspective.
However, quantum gravitational corrections could be important for objects such as Planckian quantum
black holes~\cite{CX1,CX2,CX3,CX4,CX5,Calmet:2014dea,Calmet:2011ta,Calmet:2008dg},
i.e.~hypothetical objects with a mass close to the Planck scale and size of the order of the Planck length,
which could have played an important role during the big bang.
We have seen that quantum gravity makes a difference between a static star metric and an eternal
black hole solution, the latter being described by a vacuum solution of Einstein equations.
In this section we investigate which of the two external metrics would be better suited to model a
Planckian quantum black hole.
In order to address this question, we need to extrapolate our star model into the quantum regime.
\par
In Section~\ref{S:QC} we derived quantum corrections to the metric generated by a homogeneous ball
of dust with density~\eqref{rho0} and isotropic pressure~\eqref{eq:StarPress}.
According to general relativity, this unperturbed classical configuration is
stable only provided the size of the source does not violate the Buchdahl limit~\cite{buchdahl,Wald},
so that its radius must satisfy
\be
\rs
>
\frac{9}{8}\,R_{\rm M}
\equiv
\frac{9}{8}
\left(2\,\gn\,M\right)
\ ,
\ee
where $R_M$ is the gravitational radius of the ball and would be the horizon radius of the outer
Schwarzschild metric.
While this is the classical limit, it may not hold for quantum black holes as can be seen by taking
$\rs\sim\lp\sim\sqrt{G_N}$ and $M\sim M_P\sim 1/\sqrt{G_N}$~\footnote{In this section we shall
use units with $c=1$, $\gn=\lp/M_P$ and $\hbar=\lp\,M_P$.}.
Quantum black holes are not expected to be stable objects anyway, but one expects them to decay
very quickly within a time of the order of the Planck time $\tau_P\simeq \sqrt{G_N}$.
We thus do not expect Planckian black holes to be well described by vacuum solutions.
The inside of Planckian black holes is certainly not in vacuum as the fluctuations of space-time are
expected to be large and space-time could lose its meaning altogether on such short distances.
A better approximation might thus be to describe such objects might with a quantum corrected star metric.
\par
In fact, even if we accept the general relativistic prediction that the collapsed matter giving rise
to a black hole geometry must end in a very small region of extremely high density~\footnote{It is worth
recalling that delta-like sources in general relativity are not mathematically consistent  \cite{geroch}.},
it is not {\em a priori\/} clear that the size of this region remains negligible when the black hole mass
$M$ approaches the Planck scale. 
\par
In particular, the external metric~\eqref{eq:ExtSchw} receives the quantum
corrections~\eqref{eq:MetricCorr} in the regime $|r-\rs|\gg \lp$ (as we explained in Section~\ref{S:div}).
For $r\gg\rs$, the corrected metric can therefore be written as
\be
\d s^2
=
g_{tt}\,\d t^2
+g_{rr}\,\d r^2
-r^2\,\d\Omega^2
\ ,
\label{qGext}
\ee
with
\be
g_{tt}
&\!\!\simeq\!\!&
1
-\frac{2\,\gn\,M}{r}
- \frac{\hat{\alpha} \,\hbar\,\gn^2\,M}{r^3}
\nonumber
\\
&\!\!\simeq\!\!&
1
-\frac{2\,\lp\,M}{M_P\,r}
-\frac{\hat{\alpha} \,\lp^3\,M}{M_P\,r^3}
\ ,
\label{gtt}
\ee
and
\be
g_{rr}
&\!\!\simeq\!\!&
-\left(1-\frac{2\,\gn\,M}{r}\right)^{-1}
+\frac{\hat{\beta}\,\hbar\,\gn^2\,M}{r^3}
\nonumber
\\
&\!\!\simeq\!\!&
-\left(1-\frac{2\,\lp\,M}{M_P\,r}\right)^{-1}
+\frac{\hat{\beta}\,\lp^3\,M}{M_P\,r^3}
\ ,
\label{grr}
\ee
where $\hat{\alpha}=128\,\pi\,(\alpha+\beta+3\,\gamma)$ and $\hat{\beta}=384\,\pi\,(\alpha-\gamma)$.
Note that $\hat{\alpha}>0$ for scalar and vector particles as well as for fermions and gravitons,
while $\hat{\beta}<0$ for vectors, fermions and gravitons, and can be both positive and negative
for scalars depending on the value of the non-minimal coupling $\xi$ (see Table 1).
On considering the particle content of the Standard Model and minimal coupling $\xi=0$, one would
then find $\hat{\beta}<0$. 
\par
\begin{figure}[t]
\centering
\includegraphics[width=10cm]{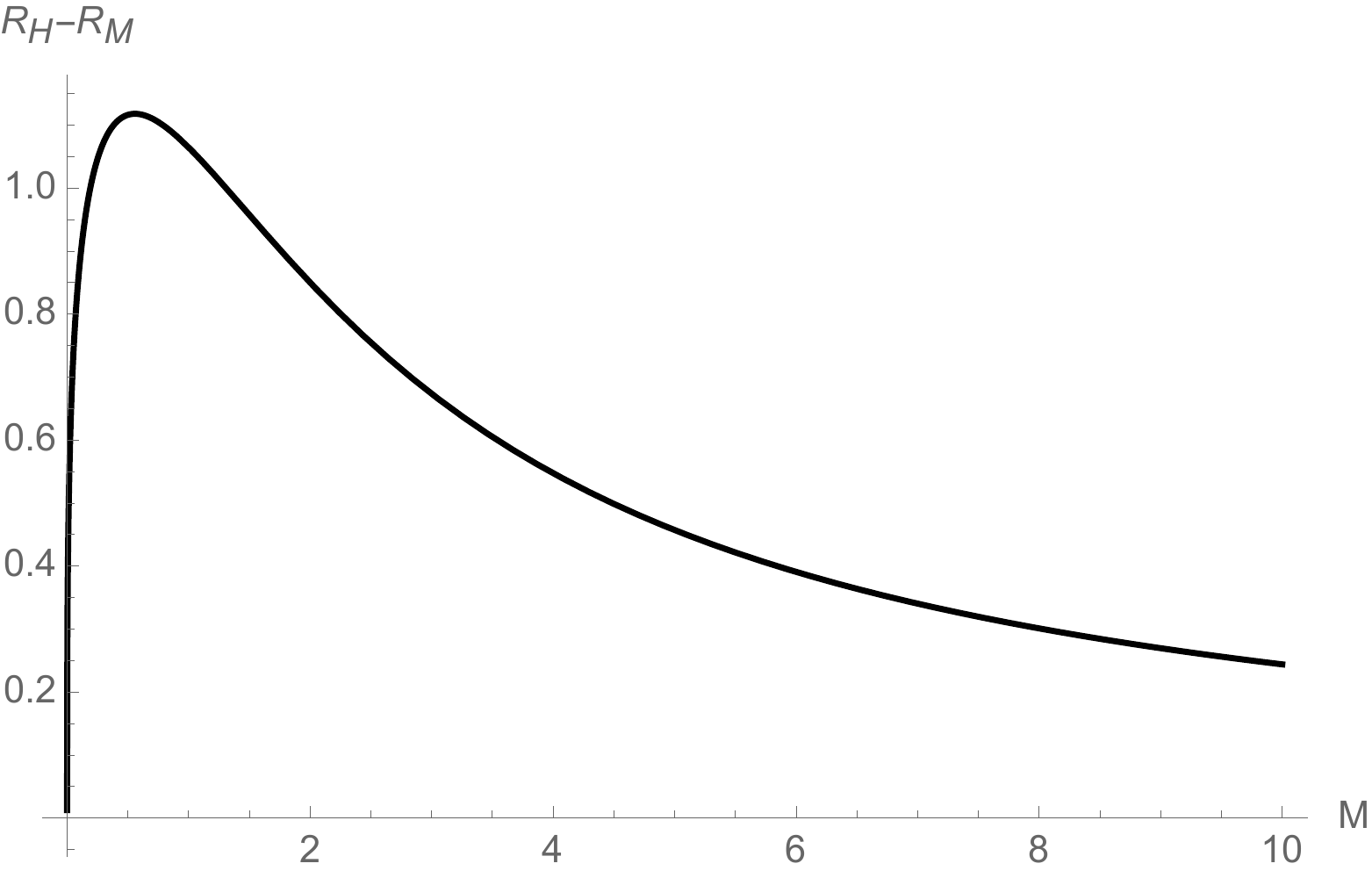}
\caption{Difference $R_H-R_M$ for $M>0$ and $\hat{\beta}=-10$ in Planck units.}
\label{Rh}
\end{figure}

The gravitational radius $\Rh$ of the system is then determined by the condition $g^{rr}(\Rh)=0$.
For $\hat\beta<0$, one finds
\begin{align}
\frac{\Rh}{\lp}
=
\frac{2\,M}{3\,M_P}
+
\left\{ -\frac{M}{2 M_P} \left[ \hat{\beta} - \frac{16 M^2}{27 M_P^2}
+
\sqrt{\hat{\beta} \left( \hat{\beta} - \frac{32 M^2}{2 7 M_P^2} \right)} \right]
\right\}^{1/3}
\nonumber
\\
+
\left\{-\frac{2M}{M_P}\left[ \hat{\beta} - \frac{16\,M^2}{27\,M_P^2} - \sqrt{\hat{\beta} \left( \hat{\beta} - \frac{32 M^2}{27 M_P^2} \right)}
\right]
\right\}^{1/3}
\ .
\label{RHq}
\end{align}
and it follows that $\Rh>R_M$ for any values of $M>0$ (see Figure~\ref{Rh}).
If we push the above description to values of the mass $M\gtrsim M_P$, 
this implies that, if the matter which sources the metric is not confined in a singularity, but
occupies a finite volume~\cite{PN} of size, say $\rs\sim\lp$, its gravitational radius is significantly larger than
it would be in the vacuum Schwarzschild geometry.
Consequently, the probability of this system of size $\rs$ to be a black hole would be larger
according to the Horizon Quantum Mechanics~\cite{hqm}.
Moreover, this is qualitatively similar to what was found in~\cite{Casadio:2015bna},
namely that the horizon area would also be larger than in general relativity.  However, one has to be careful interpreting the results obtained in Figure \ref{Rh}, since $R_H-R_M$ doesn't exceed $l_p$, which is precisely the region where our approach breaks down, as discussed in the previous section.
\par
Ideally, for sufficiently large $\hat\beta$ and small mass $M$, one could have
\be
\Rh
\gtrsim
\frac{9}{8}\,R_M
\ ,
\ee
which implies that the classical Buchdahl limit will not survive in this quantum realm as anticipated.
These considerations indicate that the metric of a Planckian quantum black hole might be better
described by our quantum corrected star model rather than by a Schwarzschild metric.
\section{Conclusions}
\setcounter{equation}{0}
\label{s:outlook}
In this paper we have calculated the full set of quantum gravitational corrections to the metric of
a star in stable equilibrium, as described by the Tolman-Oppenheimer-Volkoff equation,
to second order in curvature.
We have found a remarkable result.
While eternal black holes, which are static vacuum solutions of general relativity, and stars have
the same outside metric in general relativity, namely the famous Schwarzschild vacuum metric,
quantum gravity makes a difference between black holes and stars at second order in curvature.
Star solutions receive a quantum gravitational correction at this order, while vacuum black holes do not.
It raises a deep question, namely what happens to this correction if we were to follow the gravitational
collapse of a ball of dust?
According to our results, a distant observer would be able to monitor the collapse of the star by
measuring the quantum gravitational corrections to Newton's gravitational potential.
If he followed the process, he would have an operational procedure to determine that an eternal
black hole has formed. 
\par
It is usually argued that astrophysical black holes are well described by a Kerr metric (as they rotate),
however it is a vacuum solution and there are thus no quantum gravitational corrections
to second order in curvature. 
Our calculations thus raise deep questions about the nature of astrophysical black holes.
Are they truly vacuum solutions? 
\par
Clearly answering these questions is beyond the scope of this paper.
It would require to follow precisely quantum gravitational corrections during the dynamical
process of a star collapsing into a black hole.
\par
From a technical point of view, we have obtained an interesting result showing that the standard
textbook metric for a star~\cite{schw16,Wald} is too naive when it is assuming that matter is distributed
according to a step function at the boundary of the star.
Quantum gravity forces us to consider stars with a smooth matter profile at their surfaces. 
\par
Our results also have interesting consequences for quantum black holes.
We have argued that the quantum corrected star metric could be used as an effective metric
for a quantum black holes which, if they exist, are clearly not vacuum solutions.
\par
In conclusion, quantum gravity corrections have deep implications for black holes and stars.
Even though these corrections might be too tiny to be observable, they demonstrate that black holes
are even more mysterious than usually assumed. 
\section*{Acknowledgments}
The work of X.C. and F.K. is supported in part by the Science and Technology Facilities Council
(grant number ST/P000819/1). 
R.C. is partially supported by the INFN grant FLAG and works in the framework of activities
of the National Group of Mathematical Physics (GNFM, INdAM) and SIGRAV of Italy.
\appendix
\section{Derivation of the non-local term in equation~\eqref{gr}}
\label{A:PV}
\setcounter{equation}{0}
We will here show how to calculate the expression 
\begin{equation}
\label{expression}
\ln\!\left(\frac{\Box}{\mu^2}\right) f(t,\vec x)
\ ,
\end{equation}
for time-independent and spherically symmetric functions $f(t,\vec x)=f(r)$,
where $r=|\vec x|$.
In particular, we will consider the following two cases:
a) if $\exists\,\epsilon>0$ such that $f(r')=0$ for $|r'-r|<\epsilon$, we will find equation~\eqref{expression}
can be computed rather straightforwardly and it yields
\begin{equation}
\label{result_1}
\ln\!\left(\frac{\Box}{\mu^2}\right)\,f(r)
=
\frac{1}{r}
\int_{0}^{\infty} \left(\frac{r'}{r+r'} - \frac{r'}{|r-r'|} \right) f(r')\,\d r'
\ ;
\end{equation}
b) otherwise, if $r>0$, $f(r)\neq0$ and $\exists\,\epsilon>0$ such that $f(r')$ is smooth for $|r'-r|\leq\epsilon$,
equation~\eqref{expression} requires some care to make sense and yields
\begin{align}
\label{result_2}
\ln\!\left(\frac{\Box}{\mu^2}\right)\,f(r)
= 
\frac{1}{r}
\int_{0}^{\infty} \frac{r'}{r+r'}\, f(r')\, \d r'
-
\lim_{\epsilon\to 0^+}
&
\left\{
\frac{1}{r}
\int_{0}^{r-\epsilon} \frac{r'}{r-r'}\, f(r')\, \d r'
+
\frac{1}{r}
\int_{r+\epsilon}^{\infty} \frac{r'}{r'-r}\, f(r')\, \d r' 
\right.
\nonumber
\\
&
\left. 
\phantom{\frac{1}{r}}
+2\, f(r)\left[
\gamma_E + \ln(\mu\epsilon)
\right]
\right\}
\ ,
\end{align}
which contains a Cauchy principal value integral, as was found in   \cite{X1}.  
\par
As a first step, we use time independence to express the function $f$ in terms of its Fourier transform
$\hat f$ and write
\begin{equation}
\ln\!\left(\frac{\Box}{\mu^2}\right)\,f(\vec x)
=
\int 
\frac{\d^3 k}{(2\,\pi)^3}\, 
\ln\!\left(\frac{k^2}{\mu^2}\right)
e^{i \,\vec{k}\cdot\vec{x}}\,
\hat{f}(\vec{k})
\ ,
\end{equation}
where $k=|\vec{k}|$.
Next, we use the spherical symmetry of $f$ (and $\hat f$) and assume that $\vec{x}=(0,0,r)$
without loss of generality, so that
\begin{align}
\ln\!\left(\frac{\Box}{\mu^2}\right)\,f(r)
&
=
\frac{1}{(2\,\pi)^2}
\int_{0}^{\infty} k^2\,\d k
\int_{-1}^{+1} \d(\cos\theta)\,
\ln\!\left(\frac{k^2}{\mu^2}\right)
e^{i\, k\, r\, \cos{\phi}}\, \hat{f}(k)
\nonumber
\\	
&
=
\frac{1}{2\, \pi^2\, r} 
\int_{0}^{\infty} \d k\,
k\, \ln\!\left(\frac{k^2}{\mu^2}\right)
\sin(k\, r)\,\hat{f}(k)
\nonumber
\\	
&
=
\frac{1}{\pi^2\, r}
\int_{0}^{\infty} \d k \,
k\, \ln\!\left(\frac{k}{\mu}\right)
\sin(k\, r)\, \hat{f}(k)
\ .
\end{align}
We can now Fourier transform back to coordinate space by making use of
the relation between the Fourier and the Hankel transforms for spherically symmetric
functions in 3 dimensions, namely
\begin{equation}
k^{1/2}\, \hat{f}(k)
=
(2\,\pi)^{3/2}
\int_{0}^{\infty}
r^{3/2}\, f(r)\, J_{1/2}(k\, r)\,
\d r
\ ,
\end{equation}
where $J_{1/2}(k\, r) = \sqrt{\frac{2}{\pi\, k\, r}}\, \sin(k\, r)$.
Therefore, we obtain
\begin{align}
\ln\!\left(\frac{\Box}{\mu^2}\right)\,f(r)
&
=
\frac{4}{\pi\, r}
\int_{0}^{\infty} \d k
\int_{0}^{\infty} \d r'\,
\ln\!\left(\frac{k}{\mu}\right) \sin(k r)\, \sin(k r')\, r'\, f(r')
\nonumber
\\
&
=
\frac{1}{\pi\, r} 
\int_{0}^{\infty} \d k
\int_{0}^{\infty} \d r'
\lim_{\delta\to 0^+}
\left\{
f(r')\, r'\, \ln\!\left(\frac{k}{\mu}\right)
e^{-\delta\, k} 
\right.
\nonumber
\\
&
\left.
\phantom{\frac{1}{\pi\, r} 
\int_{0}^{\infty} \d k
\int_{0}^{\infty} \d r'
\lim_{\delta\to 0^+}}
\times
\left[
e^{i\, k\,(r-r')} + e^{-i\, k\,(r-r')}
- e^{i\, k\,(r+r')} - e^{-i\, k\,(r+r')}
\right]
\right\}
\nonumber
\\
&
=
\frac{\mu}{\pi\, r}
\int_{0}^{\infty} \d r'\,
\lim_{\delta\to 0^+}
\int_{0}^{\infty} \d q\,
f(r')\, r'\, \ln(q)\, e^{-\delta\, \mu\, q} 
\nonumber
\\
&
\phantom{\frac{1}{\pi\, r} 
\int_{0}^{\infty} \d k
\int_{0}^{\infty} \d r'
\lim_{\delta\to 0^+}}
\times
\left[
e^{i\, \mu\, q\,(r-r')} + e^{-i\, \mu\, q\,(r-r')}
- e^{i\, \mu\, q\,(r+r')} - e^{-i\, \mu\, q\,(r+r')}
\right]
\ ,
\end{align}
where we rescaled the momentum variable and swapped the limit with momentum integration in the last line.
For ${\rm Re}(\alpha)>0$, we have
\begin{equation}
\int_{0}^{\infty} \d q\,
\ln(q)\, e^{-\alpha\, q}
=
-\frac{1}{\alpha} 
\left[\gamma_E + \ln(\alpha)
\right]
,
\end{equation}
which allows us to get
\begin{align}
\ln\!\left(\frac{\Box}{\mu^2}\right)\,f(r)
=
\frac{1}{\pi\, r}
\int_{0}^{\infty} \d r'\,
f(r')\, r'\, \lim_{\delta\to 0^+}
&
\left[
\frac{\gamma_E + \ln(\mu\, R_+) + i\, \phi_+}{\delta + i\, (r+r')}
+ 
\frac{\gamma_E + \ln(\mu\, R_+) - i\, \phi_+}{\delta - i\, (r+r')}
\right.
\nonumber
\\
&
\left.
\
-
\frac{\gamma_E + \ln(\mu\, R_-) + i \,\phi_-}{\delta + i\, (r-r')}
- 
\frac{\gamma_E + \ln(\mu\, R_-) - i\, \phi_-}{\delta - i\, (r-r')}
\right] 
\ ,
\label{L4}
\end{align}
where $R_{\pm} = \sqrt{\delta^2 + (r \pm r')^2}$
and $\phi_\pm= \arctan\!\left[ ({r\pm r'})/{\delta}\right]$.
The first two terms are regular and we can take the limit $\delta\to 0$ straightforwardly,
whereas the last two terms may contain a pole at $r'=r$.
Here is where the two cases mentioned above occur:
\par
\noindent
{\it Case a):}
since $f(r') = 0$ around $r$, there is no pole in equation~\eqref{L4}, which immediately yields
the result~(\ref{result_1}).
\par
\noindent
{\it Case b):} 
for $f(r)\neq 0$ but bounded and sufficiently smooth, we can rewrite equation~\eqref{L4} as
\begin{align}
\ln\!\left(\frac{\Box}{\mu^2}\right)\,f(r)
=
&
\,
\frac{1}{r}
\int_{0}^{\infty} \d r'\,
\frac{r'\, f(r')}{r+r'}
- \lim_{\epsilon\to 0^+} \frac{1}{r}
\left\{
\int_{0}^{r-\epsilon} \d r'\,
\frac{r'\, f(r')}{|r-r'|}
+
\int_{r+\epsilon}^{\infty} \d r'\,
\frac{r'\, f(r')}{|r-r'|}
\right.
\nonumber
\\
&
+
\left.
\frac{1}{\pi}
\int_{r-\epsilon}^{r+\epsilon} \d r'\,
f(r')\, r'\,
\lim_{\delta\to 0^+}
\left[
\frac{\gamma_E + \ln(\mu\, R_-) + i\, \phi_-}{\delta + i\, (r-r')}
+\frac{\gamma_E + \ln(\mu R_-) - i\, \phi_-}{\delta - i\, (r-r')}
\right]
\right\}
\nonumber
\\
=
&
\,
\frac{1}{r}
\int_{0}^{\infty} \frac{r'}{r+r'}\, f(r')\, \d r'
-
\frac{1}{r}
\lim_{\epsilon\to 0^+}
\left[
\int_{0}^{r-\epsilon} \frac{r'}{r-r'}\, f(r')\, \d r'
+
\frac{1}{r}
\int_{r+\epsilon}^{\infty} \frac{r'}{r'-r}\, f(r')\, \d r' 
\right]
\nonumber
\\
&
+L_1
\ ,
\label{Lb}
\end{align}
where it is understood that $0<\delta<\epsilon$ before the limits are taken.
The first line in equation~\eqref{Lb} already reproduces the first line in the result~\eqref{result_2}, and we
need only compute
\be
L_1
\equiv
-
\frac{1}{\pi\,r}\,
\lim_{\epsilon\to 0^+} 
\int_{r-\epsilon}^{r+\epsilon} \d r'\,
f(r')\, r'\,
\lim_{\delta\to 0^+}
\left[
\frac{\gamma_E + \ln(\mu\, R_-) + i\, \phi_-}{\delta + i\, (r-r')}
+\frac{\gamma_E + \ln(\mu R_-) - i\, \phi_-}{\delta - i\, (r-r')}
\right]
\ .
\label{Lb2} \nonumber \\ 
\ee
By swapping the limit with the integral and defining a contour around the pole
at $r'=r$, we get
\begin{align}
L_1
=
-\frac{1}{\pi\,r}\,
\lim_{\epsilon\to 0^+}
\left\{
\lim_{\delta\to 0^+}
\int_{\pi}^{2\pi}
\right.
&
{i\,\epsilon\, e^{i \,t}\,\d t}\,
(r+\epsilon\, e^{i\, t}) \, f(r+\epsilon\, e^{i\, t})
\nonumber
\\
&
\left.
\times
\left[
\frac{\gamma_E + \ln\!\left(\mu\, \sqrt{\delta^2 + \epsilon^2 \,e^{2\, i\, t}}\right)
-i \arctan\!\left(\frac{\epsilon\, e^{i\, t}}{\delta}\right)}
{\delta - i \,\epsilon\, e^{i\, t} }
\right.
\right.
\nonumber
\\
&
\phantom{\times \, [\ }
\left.
\left.
+
\frac{\gamma_E + \ln\!\left(\mu\, \sqrt{\delta^2 + \epsilon^2 \,e^{2\, i\, t}}\right)
+i \arctan\!\left(\frac{\epsilon\, e^{i\, t}}{\delta}\right)}
{\delta + i \,\epsilon\, e^{i\, t} }
\right]
\right\}
\ .
\end{align}
We can finally use the fact that $f$ is locally smooth and Taylor expand it as
$f(r+\epsilon\, e^{i\,t})= f(r) + \mathcal{O}(\epsilon)$.
Hence,
\begin{align}
L_1
=
&
-\frac{f(r)}{\pi}\,
\lim_{\epsilon\to 0^+}
\left[
\lim_{\delta\to 0}
\int_{\pi}^{2\pi}
i\, \epsilon e^{i \,t}  \,\d t\, 
\frac{\gamma_E + \ln\!\left(\mu\, \sqrt{\delta^2 + \epsilon^2\, e^{2 \,i\, t}}\right)
- i\, \arctan\!\left(\frac{\epsilon\, e^{i\, t}}{\delta}\right)}
{\delta - i\, \epsilon\, e^{i\, t} }
+\mathcal{O}(\epsilon)\right]
\nonumber
\\
&
-\frac{f(r)}{\pi}\,
\lim_{\epsilon\to 0^+}
\left[
\lim_{\delta\to 0}
\int_{\pi}^{2\pi} i\, \epsilon e^{i \,t}  \,\d t\, 
\frac{\gamma_E + \ln\!\left(\mu\, \sqrt{\delta^2 + \epsilon^2 \,e^{2\, i\, t}}\right)
+ i\, \arctan\!\left(\frac{\epsilon e^{i\,t}}{\delta}\right)}
{\delta + i \,\epsilon\, e^{i\, t}}
+\mathcal{O}(\epsilon)\right]
\nonumber
\\
=
&
-\frac{4\,f(r)}{\pi}\,
\lim_{\epsilon\to 0^+}
\left\{
\lim_{\delta\to 0^+}\,
\arctan\!\left(\frac{\epsilon}{\delta}\right)
\left[\gamma_E + \ln\!\left(\mu\,\sqrt{\delta^2+\epsilon^2}\right)\right]
+\mathcal{O}(\epsilon)\right\}
\nonumber
\\
=
&
-2\, f(r)
\left[ \gamma_E + \ln\!\left(\mu\,\epsilon\right)\right]
\ ,
\end{align}
which completes the result presented in equation~(\ref{result_2}).
%
%
%
%
%
%

\end{document}